\documentclass[usenatbib,psfig]{mn2e}

\usepackage{epsfig}

\begin{document}


\title[Extended line emission in barred galaxies]{Discovery of kpc-scale line emission in barred galaxies, not linked to AGN or star formation.}
\author[P. A. James \& S. M. Percival]{P. A. James\thanks{E-mail:
P.A.James@ljmu.ac.uk} \& S. M. Percival\\
Astrophysics Research Institute, Liverpool John Moores University, IC2, Liverpool Science Park, 146 Brownlow Hill, Liverpool L3 5RF, UK\\
}

\date{Accepted . Received ; in original form }

\pagerange{\pageref{firstpage}--\pageref{lastpage}} \pubyear{2015}

\maketitle

\label{firstpage}

\begin{abstract}
We present an analysis of the optical line emission from nearby
barred galaxies, and in particular look at the radial range
occupied by the bar.  In many cases this region is marked by what we
term a `star formation desert', with a marked deficit of HII regions
in optical narrow-band H$\alpha$ imaging. Here we present long-slit
spectroscopy revealing that such regions do have line emission, but
that it is low-level, spatially smooth and almost ubiquitous. The 
relative strengths of
the H$\alpha$ and the spectrally adjacent [NII] lines in the regions
are completely discrepant from those associated with star formation
regions, and more closely match expectations for `LINER' regions. We
quantify the total line emission from these extended, kpc-scale regions, and
determine the spurious contribution it would make to the determined
star formation rate of these galaxies if interpreted as normal
H$\alpha$ emission. We concur with previous studies that link this
`LINER' emission to old stellar populations, e.g. post-asymptotic giant branch stars, and
propose these strongly-barred early-type spirals as a prime location
for further tests of such emission.
 
\end{abstract}


\section{Introduction}

 Bars are a common feature of spiral galaxies, with
near-IR imaging revealing that the majority have some bar structure in
their central regions, while strong bars are seen in about one-third
of spirals \citep{knap00, mari07}.  
They represent the most spectacular example of the complex interplay between stellar
and gas kinematics in disk galaxies.  They have been implicated in a wide range of evolutionary processes in galaxies, due principally to their being the strongest violation of axial symmetry in otherwise quiescent galaxies. This leads to strong gravitational torques, with consequent changes in the angular momentum of gas and stars that can drive rapid radial motion of material.  Thus, bars have been implicated in radial migration and mixing of stars and gas in the outer disk, gas supply to central black holes to fuel AGN activity, and wholesale transport of material from disks to bulges, with (presumed) global increases in global star formation rates, causing `secular' morphological evolution \citep{korm04} on timescales of a few Gyr.

Much of the debate relating bars and star formation has concerned the evidence
linking strong bars to the increases in the global star formation rate compared
with unbarred galaxies with otherwise similar properties. Some studies have
found evidence for such a link, with the early study of \citet{heck80} being confirmed by work using IRAS-based measurements of global star formation rates, e.g. \citet{hawa86}, \citet{huan96} and \citet{ague99}.
However, this paper will focus on a little-discussed aspect of bars, which is
their ability, in specific but important environments, to {\em suppress} star formation. This suppression was pointed out by \citet{tubb82}, and explored
observationally by the present authors in \cite{jame09}, where we noted several
strongly-barred galaxies that had strong dips in their radial H$\alpha$ line
emission profiles, in the radial range where their unbarred counterparts hosted 
their strongest star formation activity.  In this paper, we explore this result
in more detail using long-slit spectroscopy, and use these galaxies as probes 
of possible sources of line emission other than star formation, including
`LINER-type' emission and H$\alpha$ emission from diffuse interstellar gas (DIG).

\subsection{Effects of bars on patterns of star formation in galaxy disks}

\begin{figure}
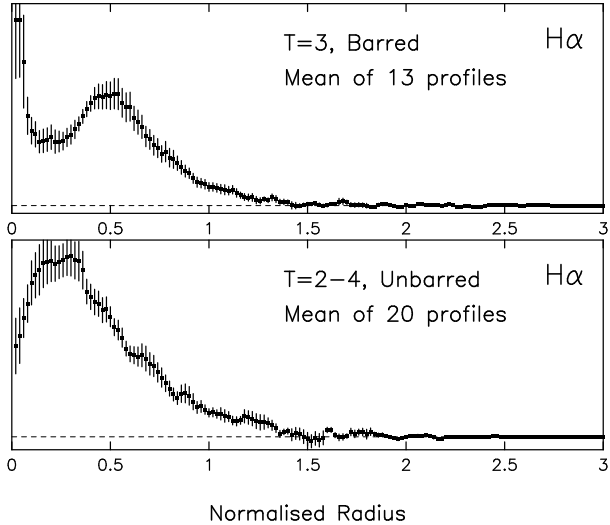

\includegraphics[height=80mm,angle=270]{prof_T3_B.ps}
\includegraphics[height=80mm,angle=270]{prof_T2_4_A.ps}
\caption{
The average radial H$\alpha$ profile for all SBb
galaxies in the H$\alpha$GS sample. Note the pronounced dip
in the SBb profiles, end that the radial range of this dip
corresponds to the peak in the line emission of unbarred
galaxies.}
\label{fig:haprof_bar}
\end{figure}

\begin{figure}
\includegraphics[width=80mm,angle=0]{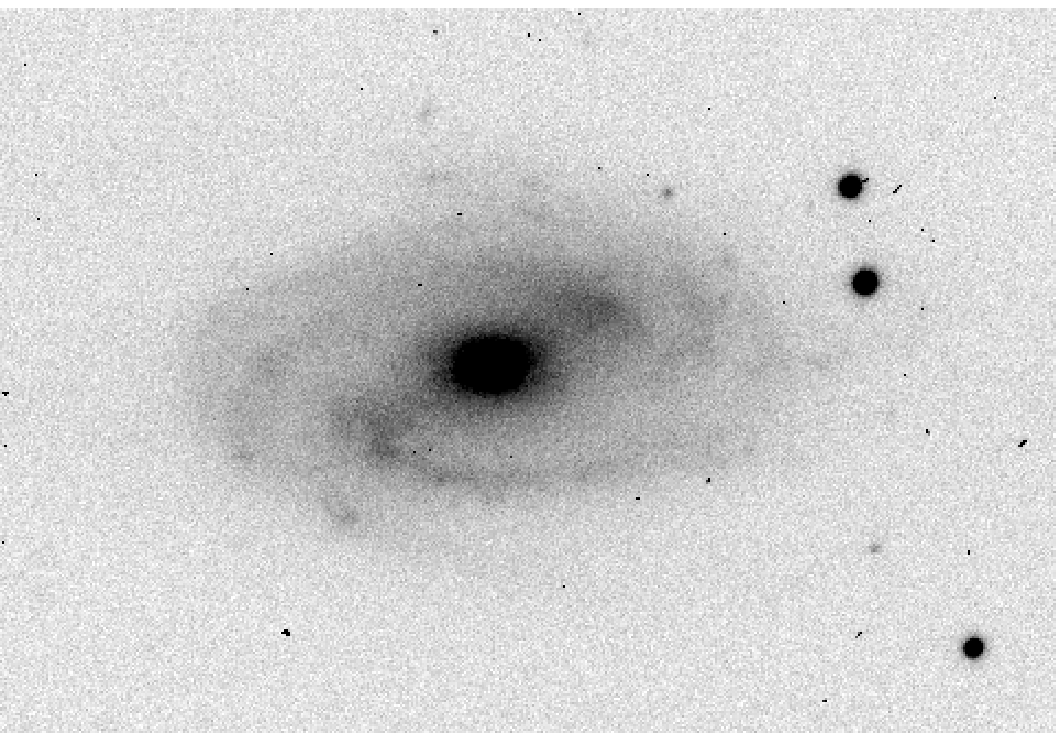}
\includegraphics[width=80mm,angle=0]{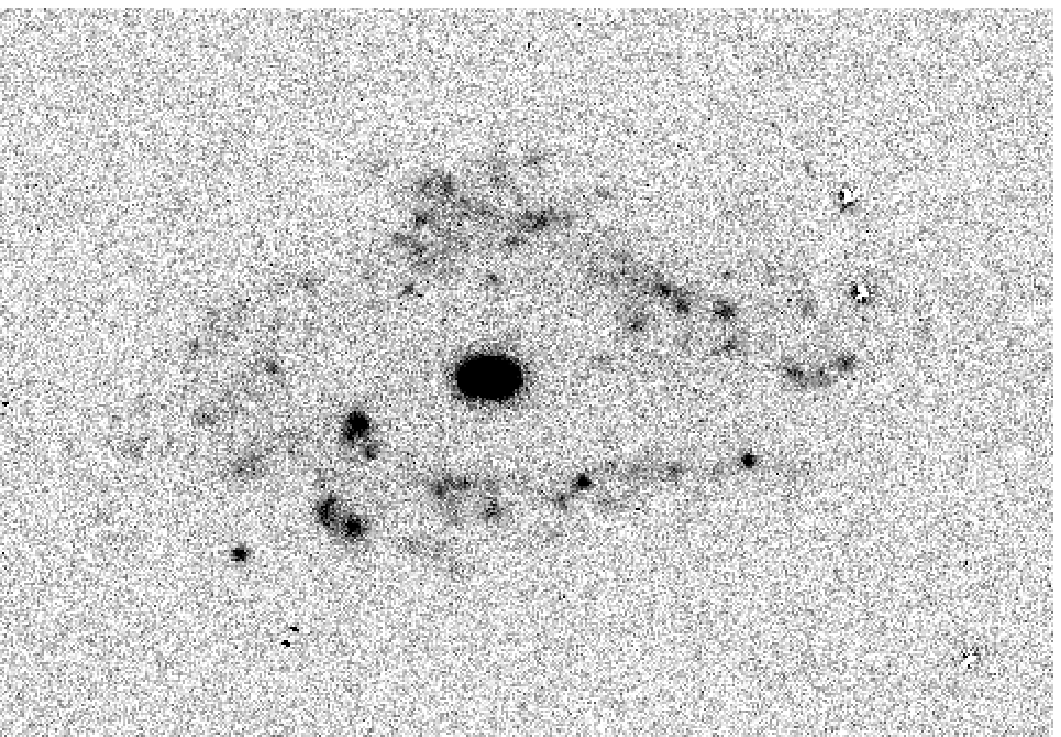}
\caption{NGC~2712, a strongly-barred SBb galaxy.  Shown are an $R$-band image (upper frame) and a continuum-subtracted H$\alpha$ image (lower frame), with the latter showing the characteristic ring-shaped star formation morphology investigated in this paper. The images cover an area of 168$^{\prime\prime}$ by 111$^{\prime\prime}$.}
\label{fig:N2712R_Ha}
\end{figure}

The initial motivation for the present study came from an
analysis of the star formation  properties of $\sim$ 200 barred and
unbarred disk galaxies \citep{jame09} using H$\alpha$ imaging
data from the H$\alpha$GS survey \citep{jame04}.  Part of this
analysis looked at the radial profiles of H$\alpha +$ [N{\sc ii}]
emission, averaged over all galaxies of a given Hubble type.  This
analysis uncovered a characteristic pattern in the SBb galaxies (and
to some extent in SBbc and SBc galaxies), as shown in
Fig.~\ref{fig:haprof_bar}. The upper frame in this figure
shows the average radial H$\alpha$ profile for {\bf all} SBb
galaxies in the H$\alpha$GS sample, with no other imposed selection criteria.  
The vertical axis is the fraction of the total galaxy-wide emission-line
flux contributed by annular regions; the horizontal axis is the radial size of each annular region, which is
normalised by the $R$-band 24~mag./sq.arcsec isophotal radius of the galaxy, R$_{24}$, prior to averaging. Note that the values are {\em not}
normalised by the area of each annulus, and hence differ from more frequently
used surface brightness profiles.
Figure \ref{fig:haprof_bar} (upper frame)
shows a characteristic dip in SF at intermediate radii.  
Figure \ref{fig:N2712R_Ha} shows an example of one of the galaxies
contributing to the mean SBb H$\alpha$ profile, where the images shown
are the $R$-band light in the upper frame, and continuum-subtracted H$\alpha$
(including the surrounding [NII] lines) in the lower. This
illustrates clearly the reason for these distinctive profiles
 in the barred galaxies; the
narrow-band H$\alpha$ image shows strong emission both in
a central peak, and within a ring lying just outside the ends of the
bar (the bar itself is evident in the $R$-band image, but not in line emission).  Between
these lies a region swept out by the bar which has greatly suppressed
line emission, but that shows significant continuum emission in the
$R$-band, which we have termed the `star formation desert', which will be
abbreviated henceforth as `SFD' where appropriate.  This is the feature
which we analyse in the present paper.

Note that this pattern of SF is not seen in unbarred early-type spiral
galaxies. This is shown in the lower frame of Fig.~\ref{fig:haprof_bar}, 
which is calculated in exactly the same way as the upper frame, but for 
galaxies classified as being unbarred.  Here all galaxies of types Sab,
Sb and Sbc have been included to give comparable numbers to the SBb profile,
but otherwise no selection has been imposed in making this unbarred profile.
The most interesting point in this comparison is that the `desert' region 
in the SBb profiles matches very closely with the radial range (0.2--0.4 times the R$_{24}$ isophotal radius) that makes the
strongest contribution to the total star-formation activity in the unbarred
galaxies.

Contrary to a popular view, the strong bulges of 
Sa-type galaxies are not elliptical-galaxy like regions, devoid of SF.
In fact, in \citet{jame09}, we showed that S0a and Sa galaxies had the 
most centrally-concentrated star formation of all spiral types, although
the overall star formation rate per unit stellar is lower than in later-type
spiral galaxies.  So, these SF desert regions provide an unusually clean
window to study the central few kpc of bright spiral galaxies, uncontaminated
by emission lines from HII regions.

Fig.~\ref{fig:haprof_bar} (upper frame) shows that the dip in the
mean emission line profile for SBb galaxies does not fall to zero 
emission-line flux, raising the question of how completely 
the bars suppress star formation; is this an artefact of averaging over
multiple galaxy profiles, is there residual star formation in some or 
all of these barred galaxies, or is there some other explanation, unrelated 
to star formation, for this residual flux in these regions?  This is one of 
several questions addressed in the present paper.

\subsection{Sources of emission lines}

In the context of the current paper, it is important to consider other
possible sources of emission lines, apart from star formation.  In the central
region, AGN-type emission can dominate, powered by activity associated with 
a central super-massive black hole.  This type of emission is generally 
quite cleanly distinguished from that associated with star formation, both
by its spatial distribution, and by the very different line ratios resulting 
from the much harder ionising spectrum from AGN, resulting in line ratio
diagnostics pioneered by \citet{bald81}.  More problematic, and interesting 
from the perspective of the current paper, is the `LINER' (for `Low Ionization Nuclear Emission-line Regions') emission, first studied in detail by \citet{heck80}. This emission has characteristically strong emission lines from low-ionisation states, including [OII]$\lambda$3727, [OI]$\lambda$6300, [NII]$\lambda$6548 \& 6584 and [SII]$\lambda$6717 \& 6731, relative to the Balmer emission lines
including H$\alpha$ $\lambda$ 6563 and H$\beta$ $\lambda$4861. \citet{heck80}
interpreted the LINERs as a low-luminosity extrapolation of AGN, and this interpretation was explored in more detail by \citet{dopi95}, who modeled the 
line emission in these systems as shock-excited gas, powered either by 
AGN jets or accretion-driven flows.  However, \citet{stas08} conclude, from
a population synthesis study of $>$10$^5$ galaxies from the Sloan Digital Sky Survey (SDSS), that the majority of
those showing LINER-type emission can be explained through purely stellar emission, on 
the assumption that these are post-starburst galaxies, termed `retired' galaxies by \citet{stas08}.

Other evidence for diffuse emission lines is found in the analysis of 11 spiral galaxies within 10~Mpc by \citet{thil02}. After excluding the emission from discrete regions of H$\alpha$ $+$ [NII] emission, typically 45\% of the total remains as a diffuse component, which they conclude is partially, but not completely, explained by the leakage of ionising photons from HII regions.

In a recent study, \citet{brem13} find LINER-type emission over the central
4~kpc region of the strongly-barred nearby SBb galaxy NGC~5850.  They 
follow \citet{stas08} in ascribing this emission to a purely stellar origin,
and particularly post-Asymptotic Giant Branch (p-AGB) stars.  However, NGC~5850
may have recently had a high-velocity encounter
with the nearby elliptical galaxy NGC~5846 \citep{higd98} raising the possibility that
shock excitation may also be involved.

Thus, the major questions addressed in the present paper are as follows:\\
\begin{itemize}
\item Are the apparent `desert' regions seen in narrow-band imaging completely
devoid of line emission?
\item If weak lines are found, are the line ratios consistent with star formation, LINER-type emission, or a combination of the two?
\item Does line emission correlate in intensity with the surface brightness of the underlying stellar population, as expected for lines resulting from an evolved stellar population (e.g. p-AGB stars)?
\item Are the velocities of any lines consistent with their coming from a rotating disk population?
\end{itemize}

The structure of the paper is as follows. The main properties of the observed galaxy sample are introduced, and the spectroscopic observations and their reduction are described in section 2.  Section 3 includes the main results in terms of emission line detections, the comparison of line ratios from `desert' and star formation regions, and a brief discussion of the kinematics of the detected emission.  Section 4 presents a discussion of the interpretation of the observations in terms of the likely origin of the line emission.  The main conclusions are then summarised in section 5.

Throughout the paper, galaxy distances are calculated assuming a Hubble constant of 68~km~s$^{-1}$~Mpc$^{-1}$ \citep{plan13} and the Virgo-infall-only model from the NASA Extragalactic Database (NED), for galaxies with recession velocities greater than 2000~km~s$^{-1}$.  Below this limit, values based on redshift-independent distance indicators (e.g. Cepheids or type Ia supernovae), also taken from NED, were used.

\section{Observations}

Observations are presented here for 15 low-redshift barred spiral
galaxies.  The sample is not in any sense claimed to be a complete or
statistically representative sample of galaxies, with the main
criterion for selection being that we have narrow-band H$\alpha$
imaging which shows evidence for a clear `desert' over a radial range
corresponding to a clearly visible bar.  Note that the evidence for a
bar comes from broad-band imaging, as bars themselves are generally
not visible in H$\alpha$ imaging, as shown in
Fig.~\ref{fig:N2712R_Ha}. The major properties of the 15 galaxies are
listed in Table \ref{tab:gals_obs}.  Ten are classified as
strongly-barred (SB) types, with the remaining 5 having SAB
classifications.  In terms of Hubble classification, 8 are Sb types,
with the remainder being spread between Sa and Scd. Column 3 of Table~\ref{tab:gals_obs}
lists any classifications or comments in NED relating to the nuclear 
properties of the galaxies; 7 have no relevant comments, 5 have been 
noted as AGN or Seyferts, and 3 have nuclear regions dominated by star 
formation activity. All of these galaxies except UGC~3685 are within the footprint of the SDSS, and seven have nuclear spectral data from SDSS enabling classifications of the dominant nuclear ionisation source using the methods of \cite{bald81}. These are listed in column 4,  where `Comp' indicates a composite spectrum with evidence for both AGN and stellar contributions to the ionising flux.
Recession velocities in km~s$^{-1}$ are given in column 5, and the adopted galaxy distances in Mpc in column 6.
All 15 galaxies in the present sample are viewed at
moderate inclinations, in the range 29$^{\circ}$ - 60$^{\circ}$, as listed in column 7. The position angle (PA) of the galaxy major axis is in column 8 and the orientation of the bar, measured from our own imaging, is in column 9.

The new observations presented here are long-slit optical spectra,
taken with the Intermediate Dispersion Spectrograph (IDS) on the
Cassegrain focus of the 2.5-metre Isaac Newton Telescope (INT) at the
Roque de Los Muchachos Observatory on La Palma in the Canary
Islands.  The observing time was allocated by the UK Panel for the Allocation of 
Telescope Time to proposal I/2014A/08.
Most of the observations presented here were taken between the 4th and 8th
February 2014, using the R1200Y grating, the RED$+$2 CCD detector and
a 1.5$^{\prime\prime}$ slit. This combination gives a dispersion of
0.53~\AA\ pixel$^{-1}$, an unvignetted wavelength range of 1166~\AA\,
and the slit width projects to an effective resolution of 1.19
\AA. The integration times were usually 3$\times$1200~s; the actual
values are listed in Table \ref{tab:gals_obs} (col. 10), which also
lists the night on which the observation was made (col. 9) and the
position angle in degrees E of N (col. 11).  The slit angle was generally
chosen so as to maximise the area of star formation desert region contained
within the slit and thus was not, in general, at the parallactic angle.
However, atmospheric refraction effects are not a problem for the present 
study since the only line ratios we are interested in are H$\alpha$ and the 
surrounding [NII] lines, which cover a very small wavelength range making
differential refraction negligible. The final column in Table \ref{tab:gals_obs}
is the outer radial limit used for the extraction of the SFD region spectra
in kpc.

Spectrophotometric standards taken from the Isaac Newton Group list 
(available at http://catserver.ing.iac.es/landscape/tn065-100/workflux.php)
were observed on all nights at a range of airmasses to enable flux calibration.
All observations were taken in clear, photometric conditions, with seeing generally between 1.2 and 2.0$^{\prime\prime}$.

\begin{table*}
 \begin{minipage}{140mm}
  \caption{Properties of the observed galaxies and details of the observations.}
  \begin{tabular}{llllrrrrrccrc}
  \hline
 Name      &  Classn.   &  Nucl.      & Nucl.  &   Vel  & Dist &  Inc     &   PA   &  Bar   &  Date obs  &  Int time  &  Slit  & R$_{\rm MAX}$\\
           &            &  NED        & SDSS   &         & Mpc &          &        &        &            &    s       &        &  kpc    \\
   \hline            
NGC~864    &  SAB(rs)c  &             &        &  1562  & 18.2 &    41    &    20  &   95   &  20140206  &   2x1200   &    55  &  3.0  \\
NGC~2268   &  SAB(r)bc  &             &        &  2222  & 37.6 &    52    &    66  &   40   &  20140204  &   3x1200   &     3  &  1.3  \\
UGC~3685   &  SB(rs)b   &             &        &  1797  & 27.9 &    34    &   135  &  131   &  20140205  &   3x1200   &     0  &  2.4  \\
NGC~2543   &  SB(s)b    &             & SF     &  2471  & 38.6 &    55    &    45  &   90   &  20140204  &   1x1200   &     0  &  4.5  \\
NGC~2595   &  SAB(rs)c  &             & Comp   &  4330  & 64.4 &    41    &    45  &  158   &  20140205  &   3x1200   &    80  &  8.3  \\
NGC~2712   &  SB(r)b    &             &        &  1815  & 31.2 &    57    &   178  &   28   &  20140207  &   3x1200   &   170  &  3.3  \\
NGC~3185   &  (R)SB(r)a &     Sy 2    & Sy 2   &  1217  & 23.4 &    47    &   130  &  113   &  20140204  &   3x1200   &   140  &  3.3  \\
NGC~3351   &  SB(r)b    &  Starburst  &        &   778  & 10.2 &    47    &    13  &  110   &  20140205  &   3x1200   &    20  &  2.5  \\
NGC~3367   &  SB(rs)c   &  Sy,LINER   &        &  3040  & 46.6 &    29    &  5-70  &   72   &  20140206  &   3x1200   &   170  &  2.3  \\
NGC~3811   &  SB(r)cd   &  Starburst  &        &  3105  & 49.8 &    41    &   160  &   30   &  20140210  &   3x1000   &    87  &  2.6  \\
NGC~4051   &  SAB(rs)bc &  Sy1.2/1.5  &        &   700  & 14.8 &    42    &   135  &  128   &  20130113  &   2x1200   &   117  &  2.3  \\
NGC~4210   &  SB(r)b    &             & LINER  &  2732  & 45.1 &    39    &   105  &   45   &  20140206  &   3x1200   &   170  &  1.3  \\
NGC~5698   &  SBb       &  HII        & SF     &  3679  & 58.8 &    60    &    70  &  168   &  20140208  &   3x1200   &    70  &  6.3  \\
NGC~5806   &  SAB(s)b   &  Sy2,AGN    & Comp   &  1359  & 25.5 &    59    &   170  &  172   &  20140207  &   3x1100   &    30  &  1.6  \\
UGC~10888  & (R')SB(r)b &   AGN       & Comp   &  6149  & 95.5 &    52    &   149  &   69   &  20140208  &   3x1000   &   120  &  4.3  \\
\hline
\end{tabular}
\label{tab:gals_obs}
\end{minipage} 
\end{table*}

\subsection{Data reduction}

Data reduction was performed using Starlink software.  Most of the reduction, including bias subtraction, flat fielding, correction of spectra for minor rotation and spatial distortion effects, and sky background subtraction, was completely standard and will not be described in detail here.  Wavelength calibration was performed using arc lamp spectra taken with the light from a copper neon$+$argon arc lamp, taken immediately before or after the galaxy spectrum and hence with the telescope in the same position, to avoid flexure errors.  The spectra were flux calibrated from the spectrophotometric standard observations from the appropriate night.

Once the reduced two-dimensional spectra had been produced (see an example in Fig.~\ref{fig:u9645_2dspec}), the next task was to decide on the region of slit from which the star formation desert
spectra were to be extracted.  This process was aided by our having continuum-subtracted narrow-band H$\alpha$ imaging for all of our target
galaxies (see Fig.~\ref{fig:N2712R_Ha}, lower frame and Fig.~\ref{fig:ugc4273_slit} for examples), which were inspected with a superposed line at the 
observed position angle. This slit position is illustrated by the vertical line in Fig.~\ref{fig:ugc4273_slit}, which also highlights in white
the two parts of the slit used to extract desert regions spectra. The nuclear regions of this galaxy, NGC~2543, are strongly emitting and must
obviously be carefully excluded from the desert region analysis; this is always the case in the galaxies in the present study, giving two
desert regions for each, one on either side of the nucleus. However, the large size of one of the galaxies, NGC~4051, meant that the slit was offset
to include just the nucleus and one half of the disk, to ensure that the slit reached a `sky' area off the edge of the disk, and hence only 
one desert region was observed. (Note also that NGC~4051 was observed in an earlier observing run than that for all other data presented here, with a slightly different observational set-up.) The outer limits of the extracted desert area were defined by inspection of the narrow band images, again 
as shown in Fig.~\ref{fig:ugc4273_slit}.  The result of this process was 29 spectra from the 15 galaxies listed in Table \ref{tab:gals_obs}; these
spectra are listed in Table \ref{tab:EW_ratios_slits}.  In addition, a region of strong line emission, corresponding to a star formation region 
lying outside the desert region, and not in the nucleus, was also defined for each galaxy.  These gave us reference regions from within the same galaxy 
sample with which to compare the line ratios measured for the star formation desert regions.  The reference region spectral properties are also listed
in  Table \ref{tab:EW_ratios_slits}, and identified as `HII' in the second column.

The main properties that were extracted from these spectra were line fluxes, line widths and central wavelengths for the H$\alpha$ 6563\AA\ line, and the same parameters for the stronger of the two [NII] lines at 6584\AA. The line-fitting process also yielded a true continuum value, enabling equivalent widths to be calculated for both lines.  The one complicated case was a Seyfert galaxy, NGC~3367, where the SFD regions showed narrow lines sitting on an extremely broad emission line, where the latter is presumably associated with the central AGN.  Here, the line fluxes for the narrow H$\alpha$ and [NII] lines were calculated relative to a very local continuum level, effectively removing the contribution of the broad line emission.  In addition, a true stellar continuum level was also measured from spectral regions flanking the broad line, for use in later analysis where we investigate correlations between line strengths and the stellar continuum level.

\begin{table*}
 \centering
 \begin{minipage}{140mm}
  \caption{Observed regions, line equivalent widths and line ratios.}
  \begin{tabular}{llrcrccccccl}
  \hline
    Name     &  Region  &   EW$_{\rm H\alpha}$   &   Err   &  EW$_{\rm [NII]}$  &   Err   &  [NII]/H$\alpha$  &  [NII]/H$\alpha_C$  &  Err  &  Radial range ($^{\prime\prime}$) &  Direction \\
  \hline            
 NGC~864   &    SFD1  &    ---   &   ---   &   0.88  &   0.16  &   $>$1   &  0.801  &  0.214  &  10.12-32.12  &  NE     \\
 NGC~864   &    SFD2  &    1.31  &   0.16  &   ---   &   ---   &   0.000  &  0.000  &   ---   &  19.36-33.88  &  SW	    \\
 NGC~864   &    HII   &   93.18  &   0.93  &  20.81  &   0.84  &   0.223  &  0.221  &  0.009  &  67.76-73.92  &  NE	    \\
 NGC~2268  &    SFD1  &    1.94  &   0.12  &   2.46  &   0.16  &   1.267  &  0.808  &  0.083  &   3.08-4.84   &  N	    \\
 NGC~2268  &    SFD2  &    2.81  &   0.09  &   2.85  &   0.13  &   1.017  &  0.730  &  0.055  &   3.96-7.04   &  S	    \\
 NGC~2268  &    HII   &   72.14  &   1.03  &  25.32  &   0.51  &   0.351  &  0.346  &  0.009       &   8.36-11.44  &  S	    \\
 UGC 3685  &    SFD1  &    0.83  &   0.15  &   1.72  &   0.28  &   2.076  &  0.889  &  0.186  &   7.48-17.60  &  N	    \\
 UGC 3685  &    SFD2  &    1.60  &   0.13  &   1.44  &   0.21  &   0.895  &  0.530  &  0.092  &   7.48-17.60  &  S	    \\
 UGC 3685  &    HII   &   48.09  &   0.69  &  13.63  &   0.45  &   0.283  &  0.277  &  0.010  &  30.80-36.08  &  S	    \\
 NGC~2543  &    SFD1  &    1.19  &   0.21  &   1.17  &   0.31  &   0.983  &  0.509  &  0.149  &   7.48-24.20  &  N	    \\
 NGC~2543  &    SFD2  &    1.67  &   0.20  &   1.88  &   0.37  &   1.125  &  0.677  &  0.151  &   6.60-19.80  &  S	    \\
 NGC~2543  &    HII   &   86.41  &   1.73  &  26.75  &   0.77  &   0.310  &  0.306  &  0.011  &  23.32-26.40  &  S	    \\
 NGC~2595  &    SFD1  &    ---   &   ---   &   ---   &   ---   &    NA    &  0.000  &   ---   &   9.68-19.36  &  E	    \\
 NGC~2595  &    SFD2  &    2.59  &   0.29  &   1.79  &   0.30  &   0.693  &  0.486  &  0.094  &  13.20-26.40  &  W	    \\
 NGC~2595  &    HII   &   26.31  &   0.34  &  10.72  &   0.25  &   0.408  &  0.391  &  0.011  &  23.76-29.92  &  E	    \\
 NGC~2712  &    SFD1  &    1.00  &   0.11  &   1.44  &   0.16  &   1.441  &  0.686  &  0.107  &   6.16-21.56  &  S	    \\
 NGC~2712  &    SFD2  &    1.49  &   0.10  &   1.59  &   0.15  &   1.069  &  0.614  &  0.082  &   7.48-20.68  &  N	    \\
 NGC~2712  &    HII   &   19.66  &   0.33  &   7.58  &   0.23  &   0.385  &  0.365  &  0.013  &  33.00-44.44  &  N	    \\
 NGC~3185  &    SFD1  &    0.39  &   0.13  &   1.14  &   0.11  &   2.908  &  0.763  &  0.149  &   9.68-21.56  &  SE	    \\
 NGC~3185  &    SFD2  &    0.45  &   0.11  &   0.84  &   0.12  &   1.862  &  0.538  &  0.112  &  11.44-29.04  &  NW	    \\
 NGC~3185  &    HII   &   10.83  &   0.21  &   4.34  &   0.18  &   0.401  &  0.364  &  0.018  &  33.44-42.68  &  NW	    \\
 NGC~3351  &    SFD1  &    ---   &   ---   &   1.95  &   0.42  &   $>$1   &  1.767  &  0.512  &  27.28-51.48  &  NNE    \\
 NGC~3351  &    SFD2  &    ---   &   ---   &   1.33  &   0.13  &   $>$1   &  1.208  &  0.261  &  25.52-40.92  &  SSW    \\
 NGC~3351  &    HII   &   16.37  &   0.40  &   4.99  &   0.35  &   0.305  &  0.285  &  0.022  &  73.48-58.08  &  NNE    \\
 NGC~3367  &    SFD1  &    2.71  &   0.25  &   3.45  &   0.26  &   1.271  &  0.904  &  0.071  &   5.28-9.68   &  N	    \\
 NGC~3367  &    SFD2  &    3.21  &   0.24  &   3.94  &   0.28  &   1.226  &  0.913  &  0.094  &   5.72-10.12  &  S	    \\
 NGC~3367  &    HII   &   46.25  &   0.58  &  13.55  &   0.25  &   0.293  &  0.286  &  0.006  &  11.88-16.28  &  N	    \\
 NGC~3811  &    SFD1  &    1.13  &   0.11  &   1.07  &   0.13  &   0.950  &  0.481  &  0.078  &   5.28-10.56  &  W	    \\
 NGC~3811  &    SFD2  &    0.35  &   0.10  &   1.24  &   0.13  &   3.581  &  0.857  &  0.163  &   5.28-10.56  &  E	    \\
 NGC~3811  &    HII   &  103.17  &   2.43  &  33.05  &   0.92  &   0.320  &  0.317  &  0.012  &  21.56-24.64  &  W	    \\
 NGC~4051  &    SFD   &    0.19  &   0.04  &   0.56  &   0.04  &   2.897  &  0.723  &  0.125  &  12.76-32.12  &  NW	    \\
 NGC~4051  &    HII   &   34.88  &   0.92  &  11.49  &   0.35  &   0.329  &  0.324  &  0.013  &  71.28-81.84  &  NW	    \\
 NGC~4210  &    SFD1  &    0.65  &   0.11  &   1.30  &   0.11  &   2.017  &  0.745  &  0.119  &   3.08-5.72   &  N	    \\
 NGC~4210  &    SFD2  &    ---   &   ---   &   1.07  &   0.14  &   $>$1   &  0.967  &  0.226  &   3.52-6.16   &  S	    \\
 NGC~4210  &    HII   &   15.99  &   0.08  &   6.15  &   0.20  &   0.385  &  0.360  &  0.013  &  11.00-15.40  &  N	    \\
 NGC~5698  &    SFD1  &    1.05  &   0.25  &   1.93  &   0.24  &   1.831  &  0.894  &  0.176  &   6.16-22.00  &  WSW    \\
 NGC~5698  &    SFD2  &    ---   &   ---   &   2.64  &   0.20  &   $>$1   &  2.394  &  0.494  &   7.04-19.36  &  ENE    \\
 NGC~5698  &    HII   &   74.92  &   0.63  &  24.39  &   0.56  &   0.326  &  0.321  &  0.008  &  33.88-37.84  &  ENE    \\
 NGC~5806  &    SFD1  &    0.84  &   0.07  &   1.74  &   0.13  &   2.072  &  0.895  &  0.123  &   6.60-13.20  &  SW	    \\
 NGC~5806  &    SFD2  &    0.60  &   0.09  &   1.65  &   0.12  &   2.744  &  0.967  &  0.150  &   6.60-11.44  &  NE	    \\
 NGC~5806  &    HII   &    6.65  &   0.16  &   2.92  &   0.14  &   0.440  &  0.378  &  0.022  &  15.40-24.20  &  NE	    \\
UGC~10888  &    SFD1  &    0.89  &   0.16  &   2.08  &   0.38  &   2.332  &  1.044  &  0.237  &   3.96-9.24   &  NW	    \\
UGC~10888  &    SFD2  &    ---   &   ---   &   0.92  &   0.15  &   $>$1   &  0.837  &  0.213  &   3.96-8.80   &  SE	    \\
UGC~10888  &    HII   &   22.08  &   0.32  &   9.42  &   0.29  &   0.427  &  0.406  &  0.014  &  10.56-18.48  &  NW	    \\
\hline
\end{tabular}
\label{tab:EW_ratios_slits}
\end{minipage} 
\end{table*}

\section{Results}

\begin{figure}
\includegraphics[width=85mm,angle=0]{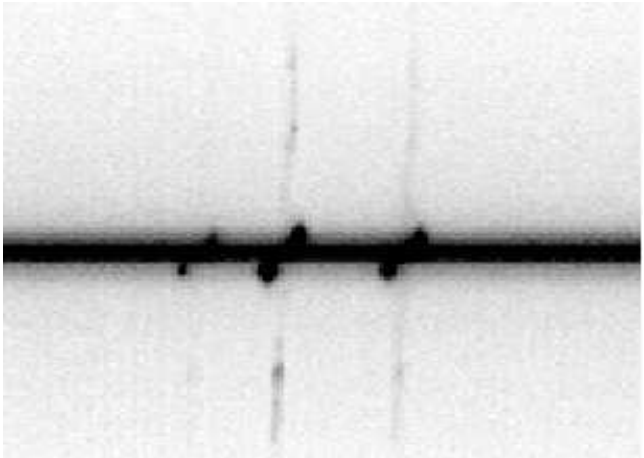}
\caption{Two-dimensional spectrum of barred galaxy NGC~5806. The vertical
extent shown corresponds to 66$^{\prime\prime}$ along the slit, centred on the
galaxy nucleus.}
\label{fig:u9645_2dspec}
\end{figure}

\begin{figure}
\includegraphics[width=85mm,angle=0]{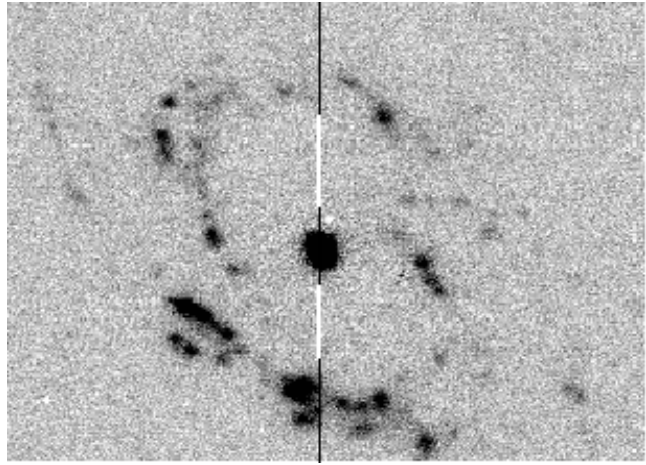}
\caption{Barred galaxy NGC~2543, showing the slit orientation
used for our IDS spectroscopy, with the extracted SFD regions
highlighted in white. The image covers an area of 120$^{\prime\prime}$ 
by 87$^{\prime\prime}$.
}
\label{fig:ugc4273_slit}
\end{figure}

The first result that is apparent from Table \ref{tab:EW_ratios_slits} is that the desert 
regions are, in fact, far from devoid of line emission.  Of the 29 regions observed, only one,
SFD1 in NGC~2595, had no detectable flux at expected locations of either H$\alpha$ or of the 
stronger of the [NII] lines at 6584~\AA. (References to [NII] detections, fluxes, and equivalent
widths henceforth will always refer to this line.) One additional region was not detected in the [NII] line only, 
and 6 gave no detectable H$\alpha$ emission.  The remaining 21 regions had detections of both lines, at a 
minimum of a 3-$\sigma$ level. The emission is typically diffuse and extended, at least at the spatial resolution of our spectral data, as shown for one representative case in Fig.~\ref{fig:u9645_2dspec}. As can be seen from the final column of Table \ref{tab:gals_obs},
the SFD region observations extend out to several kpc from the galaxy nuclei, so this is genuinely kpc-scale 
diffuse emission.

\begin{figure}
\includegraphics[width=70mm,angle=270]{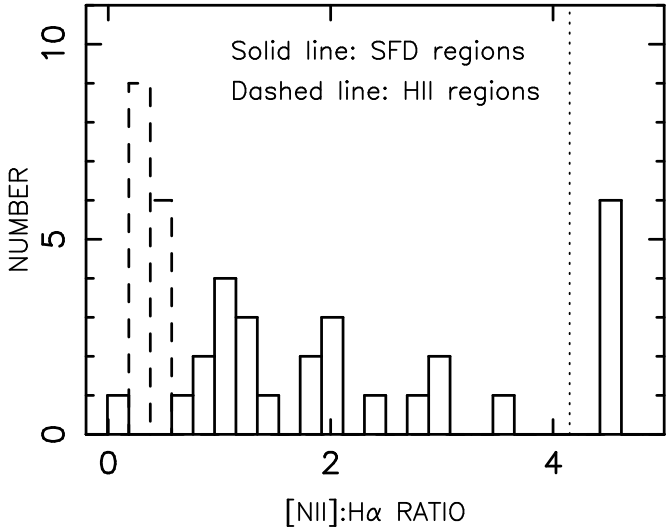}
\caption{Histogram of [NII]:H$\alpha$ line ratios for SFD (solid line)
and HII (dashed line) regions within our barred galaxy sample.
The H$\alpha$ fluxes have no correction for underlying H$\alpha$
absorption features. The points to the right of the dotted line correspond
to regions detected in [NII] only.}
\label{fig:sfdhist}
\end{figure}

Fig.~\ref{fig:sfdhist} shows the distribution of values of the observed
[NII]/H$\alpha$ line ratio.   Ratios derived from SFD regions are shown as the solid
histogram, while the corresponding values for the comparison HII regions are shown
using dashed lines.
For the latter, the ratios form a very tight distribution 
around the values typical of excitation by star formation as found by numerous 
authors, e.g. \citet{bald81}.  For the 15 HII regions in the present sample, the mean 
observed [NII]/H$\alpha$
ratio is 0.346$\pm$0.016, median 0.330.  For the SFD regions, 
the ratios are much larger on average, and for 
6 of these regions, only [NII] is detected. The latter are plotted in Fig.~\ref{fig:sfdhist}
to the right of the dotted line, at an arbitrary value for the line ratio  of 4.5; in fact, only lower limits 
can be placed for these regions.  
There is additionally one SFD region in which a detection was 
made of H$\alpha$ but not [NII], and in NGC~2595, as noted above, neither line was 
detected.\\

Neglecting the
7 regions for which the H$\alpha$ line was not detected leaves 22 SFD regions for which we can derive
emission line ratios.  For these 22, the mean observed [NII]/H$\alpha$ ratio is 1.65$\pm$0.19, median
1.36.  A Kolmogorov-Smirnov (K-S) test gives a maximum $D$ value of
0.9545 for the SFD and HII region distributions, which strongly excludes the possibility 
(formal probability $P=$0.000 to three significant figures) that the SFD and HII region 
values are 
selected from the same parent distribution. The 6 regions with [NII] line detections and
no H$\alpha$ only add weight to this conclusion.

\subsection{Correction for underlying H$\alpha$ absorption}

One possible explanation of the relatively weak H$\alpha$ emission lines
is that the measured flux is reduced due to the presence of H$\alpha$ absorption
features.  This is a particular problem in the case of this `star formation desert' analysis due to the low equivalent width of the detected emission features, which can potentially be much more strongly affected by absorption than is the case for strong lines from star formation regions, in which context H$\alpha$ absorption is often neglected as being unimportant compared with, for example, the effects of extinction.  However, it is not trivial to correct for such absorption effects.  In their pioneering study of LINER emission, \cite{ho93} subtracted the spectrum of a non-star-forming template galaxy to correct for the underlying continuum shape, but the ubiquity of the line emission found in the present sample must cast doubt on whether any observed stellar population can safely concluded to be free of emission lines.  Thus, in the present study, we instead use simulated, theoretical stellar populations, with suitably conservative assumptions about the uncertainties inherent in such modelling.   

\begin{figure}
\includegraphics[width=70mm,angle=0]{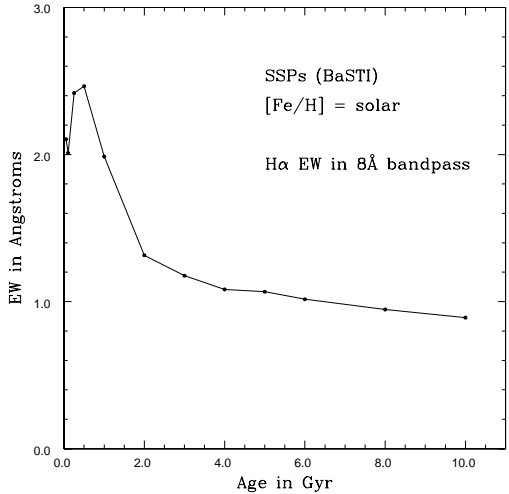}
\caption{Equivalent width of the modelled H$\alpha$ absorption line
as a function of the age of a simple stellar population.  The line is
measured in the same 8~\AA\ bandpass and continuum wavelength range used 
to determine emission line strength in the present study.
}
\label{fig:Halpha_age}
\end{figure}

We predict the underlying absorption features using simple 
stellar population (SSP) models from the BaSTI database \citep{perc09}, which
were used to produce simulated spectra at a resolution of 1~\AA~pixel$^{-1}$
for population ages from 50~Myr to 10~Gyr.
The equivalent width of the H$\alpha$ absorption line was measured in each of the resulting spectra, using the same 8~\AA\ bandpass and continuum ranges as was used to measure H$\alpha$ emission line fluxes in the observed galaxy spectra.
The resulting equivalent widths are plotted in Fig.~\ref{fig:Halpha_age}, which shows the well-established peak in Balmer line strengths for intermediate-age stellar populations; in these models and passband definitions, the highest equivalent width is 2.465~\AA\ for a 500~Myr. For ages greater than 2~Myr, the equivalent width is approximately stable at $\sim$1~\AA.

\begin{figure}
\includegraphics[width=70mm,angle=0]{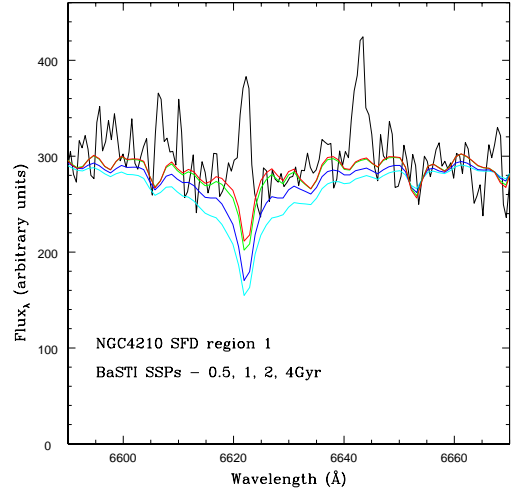}
\caption{Spectrum of the region SFD1 in NGC~4210, around the wavelength of
the H$\alpha$ line.  This is the `desert' region that shows the strongest 
evidence for H$\alpha$ absorption, from a generally depressed continuum 
level, but even in this case the modelled absorption from a 0.5 - 1 Gyr 
stellar population over-predicts the observed absorption, which is more
consistent with the 2 or 4 Gyr old simple stellar populations.
}
\label{fig:ngc4210_Ha_sfd1}
\end{figure}

On inspection of the BaSTI SSP spectra, it was clear that the younger models 
with strong H$\alpha$ features showed strong wings to their absorption lines which were not apparent in the observed spectra.  Figure~\ref{fig:ngc4210_Ha_sfd1} shows an example of one of the few observed spectra with some evidence for a
broad absorption feature around the H$\alpha$ emission line (region SFD1 in NGC~4210).  Even in this case, it is apparent that the 0.5 and 1~Gyr SSP spectra significantly over-estimate
the strength of the absorption feature that is present in this spectrum, while 2 and 4~Gyr populations reproduce its strength reasonably well.  Given that this is a conservative estimate, with all other observed spectra showing even less
evidence for absorption than is seen in region NGC~4210 SFD1, we conclude that the underlying spectra could be represented by SSPs with ages anywhere between 2 and 10~Gyr.  This corresponds to an H$\alpha$ equivalent width between 1.315 and 0.891~\AA, with an average value of 1.103$\pm$0.21~\AA.  We adopt these values as the most likely value, and associated uncertainty, of the H$\alpha$ absorption equivalent width from the underlying stellar population. This corresponds to an SSP of age$\sim$3.5~Gyr. An equivalent width of 2.465~\AA, corresponding to the 0.5~Gyr population, can be taken as a physical upper limit to any absorption correction, although we note again that such a strong feature should be easily seen in our spectra, and is never apparent.

\begin{figure}
\includegraphics[width=70mm,angle=270]{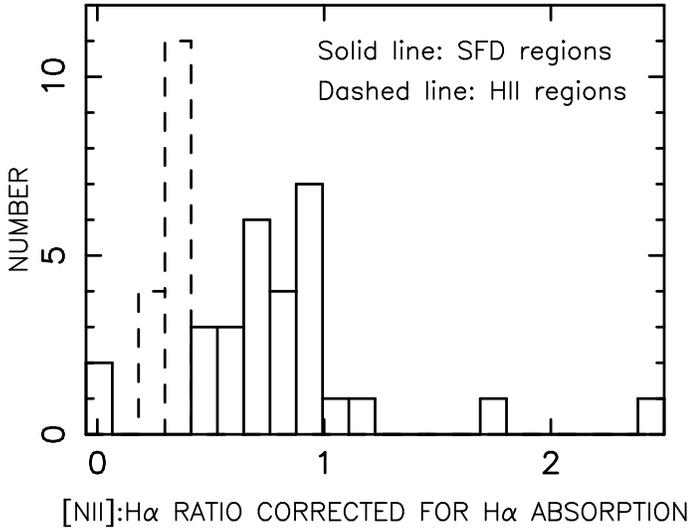}
\caption{Histogram of [NII]:H$\alpha$ line ratios for SFD (solid line)
and HII (dashed line) regions within our barred galaxy sample.
The H$\alpha$ fluxes have been corrected for the H$\alpha$ absorption 
corresponding to an old (3--4~Gyr) underlying stellar population.}
\label{fig:sfdhist_abscor}
\end{figure}

Figure~\ref{fig:sfdhist_abscor} shows the effect on [NII]/H$\alpha$ line ratios of boosting the H$\alpha$ 
fluxes by an amount corresponding to the 1.103~\AA\ equivalent width correction.  This correction has only a small effect on the HII region ratios, again shown by the dashed histogram.  For the star formation desert regions, the main effect is to tighten up the distribution with most of the regions detected in both lines having ratios between 0.4 and 1.0.  There are some additional regions contributing to Fig.~\ref{fig:sfdhist_abscor} compared to Fig.~\ref{fig:sfdhist} as we have now included regions previously undetected in H$\alpha$ line emission.  These are now given emission line fluxes corresponding to the absorption correction. This adds in one more point with a ratio of 0.0 (previously undetected in both lines, now considered just an H$\alpha$ detection), and gives defined ratios for the 6 regions plotted as lower limits to the right of the dotted line in  Fig.~\ref{fig:sfdhist}.  All 4 ratios higher than 1.0 in Fig.~\ref{fig:sfdhist_abscor} correspond to regions previously undetected in H$\alpha$.

A K-S test performed on the emission line ratios after correction for 1.103~\AA\ of underlying H$\alpha$ absorption shows that the distributions for HII and star formation desert regions are still very different.  
The 15 HII regions now have a mean [NII]/H$\alpha$ ratio of 0.330$\pm$0.013, with a median value of 0.325.  For 29 star formation desert regions, the mean ratio is 0.814$\pm$0.083 with a median of 0.801. The K-S test now finds a maximum fractional difference in the two cumulative distributions D$=$0.931, which still indicates a vanishingly small probability of their being drawn from the same parent distribution: P$=$0.000, to 3 significant figures.

For completeness, we performed the same test using a maximal absorption correction of 2.465~\AA, even though such a strong effect is inconsistent with the observed spectra.
The 15 HII regions have a mean [NII]/H$\alpha$ ratio of 0.314$\pm$0.011, median 0.320, after this correction.  For 29 star formation desert regions, the mean ratio becomes 0.458$\pm$0.039, median 0.433.  The K-S D value is reduced to 0.694, but the distributions are still formally different with the P value remaining 0.000 to 3 significant figures.

\begin{figure}
\includegraphics[width=65mm,angle=270]{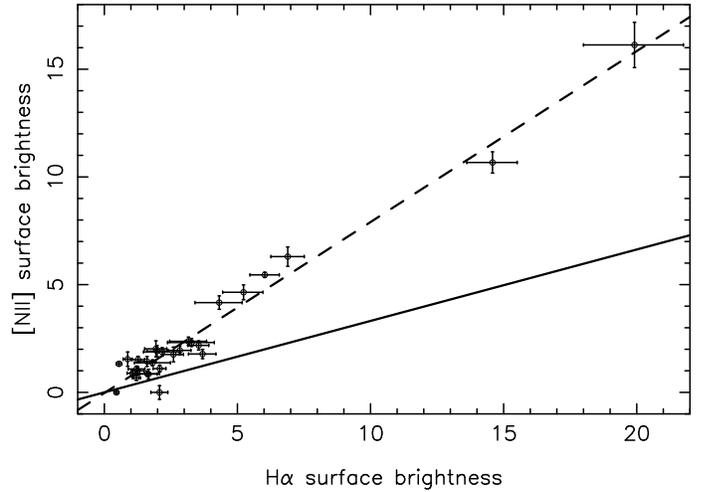}
\caption{[NII] surface brightness vs H$\alpha$ surface brightness
within star formation desert regions,
both measured in units of 10$^{-20}$W m$^{-2}$ sq. arcsec$^{-1}$.
H$\alpha$ emission has been corrected for underlying absorption
by a 3-4~Gyr old stellar population.  The solid line corresponds
to the mean emission line ratio ([NII]/H$\alpha$ = 0.332) found 
for HII regions in the same galaxy sample.}
\label{fig:nii_ha_sb}
\end{figure}

An alternative way of demonstrating the different emission-line properties
of SFD and HII regions is shown in Fig.~\ref{fig:nii_ha_sb}.  Here the [NII] emission-line surface brightness of the 29 SFD regions is plotted against their H$\alpha$ surface brightness, where the latter has been corrected for an underlying absorption line with an equivalent width of 1.103~\AA, as was used in constructing Fig.~\ref{fig:sfdhist_abscor}.  The dashed line is the regression relation of the plotted $y$ values on $x$ values; this regression line has a slope of 0.794, a $y$ intercept of --0.030, and a Pearson correlation coefficient of 0.983.  The Spearman rank correlation coefficient, which is less sensitive to extreme outliers, has a value of 0.835, confirming that the values are significantly correlated. The solid line in Fig.~\ref{fig:nii_ha_sb} corresponds to the mean line ratio found for the 15 HII regions in the same sample of galaxies, i.e. $SB_{\rm{NII}} = 0.332 \times SB_{\rm{H\alpha}}$.  Note that the points corresponding to individual HII regions cannot be plotted on Fig.~\ref{fig:nii_ha_sb} as their line surface brightnesses are orders of magnitude higher than those of any of the SFD regions.

Figure~\ref{fig:nii_ha_sb} shows clearly that the H$\alpha$ and [NII] line strengths are strongly correlated in the SFD regions, and that most of the scatter and all the strong outliers seen in the line ratios plotted in Fig.~\ref{fig:sfdhist_abscor} are due the regions with the weakest line emission, which is barely detected in some cases.  The strong correlation between the line strengths at least implies a common source driving the emission in the two lines.  However, it also appears very unlikely that this source is star formation, e.g. from a population of low-luminosity, individually undetected HII regions, as the ratio of [NII] to H$\alpha$ line strengths is significantly higher in virtually all of the SFD regions than the very narrow range of ratios found for HII regions.  Thus other sources must be sought for the ionising radiation in these regions.

\subsection{Rotational velocities}

In order to test whether this emission is closely linked to the
underlying stellar population in galaxy disks, it is useful to test
whether the line emission from the desert regions shares the
rotational velocity of other disk components.  This might not be the
case, for example, were the emitting gas to be located in galaxy
haloes.  We selected the galaxy NGC~2712 (shown in
Fig.~\ref{fig:N2712R_Ha}) for this test as it is one of the more
significantly inclined galaxies in the present study, and the slit
angle differs by only 8$^{\circ}$ from the galaxy major axis, with both
factors maximising the rotation signal.  For this test, five
velocities were calculated from different regions of our long slit
spectrum, corresponding to the nucleus, to the star formation desert
regions to the north and south of the nucleus (SFD1 and SFD2
respectively), and to star formation regions lying just beyond each of
the desert regions.  Velocities were calculated from the centroids of
gaussian profiles fitted to the H$\alpha$ lines from each of the above
regions. The resulting velocities are shown in Table~\ref{tab:rot_vel}. 

\begin{table*}
 \centering
 \begin{minipage}{140mm}
  \caption{Rotational velocities of regions within NGC~2712.}
  \begin{tabular}{lccc}
  \hline
 Region    &  Wavelength  &     Velocity     &  $\delta$V~  \\
           &     \AA      &    km~s$^{-1}$   &  km~s$^{-1}$  \\
   \hline            
Nucleus    & 6602.53      &      1815.0     &      0.0      \\
SFD1       & 6599.88      &      1693.8     & $-$121.2      \\
HII1       & 6599.40      &      1672.0     & $-$143.0      \\
SFD2       & 6605.60      &      1955.3     & $+$140.3      \\
HII2       & 6605.23      &      1938.4     & $+$123.4      \\
\hline
\end{tabular}
\label{tab:rot_vel}
\end{minipage} 
\end{table*}

This analysis makes clear that the line emission from the two
star formation desert regions shares fully in the rotational
velocity of the disk of NGC~2712.  In one case, SFD1 to the north 
of the nucleus, the rotation is $\sim$20~km~s$^{-1}$ slower than 
that of the HII region lying just outside it; but in SFD2, south of
the nucleus, the pattern is reversed, and in any case most
rotation curves show fluctuations of at least $\pm$20~km~s$^{-1}$.
So it is clear that whatever drives the SFD line emission, it is
linked to the rotating disk component.

\subsection{Diffuse line emission and the old stellar population}

\begin{figure}
\includegraphics[width=65mm,angle=270]{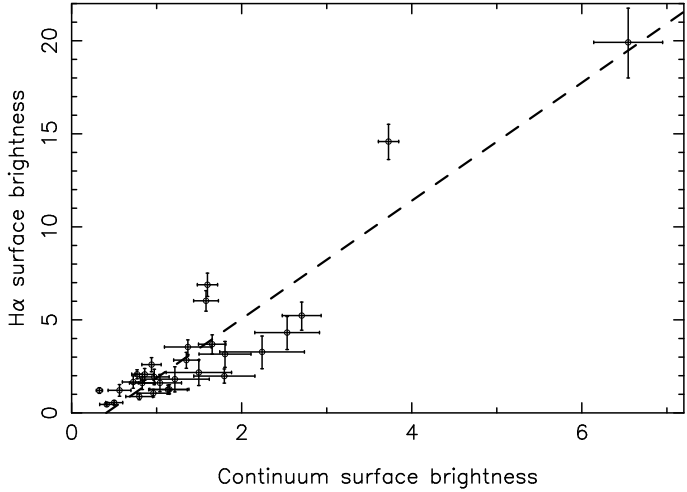}
\caption{H$\alpha$ line surface brightness vs red continuum surface
brightness
within star formation desert regions,
measured in units of 10$^{-20}$~W~m$^{-2}$ sq. arcsec$^{-1}$
and 10$^{-20}$~W m$^{-2}$ \AA $^{-1}$ sq. arcsec$^{-1}$
respectively.}
\label{fig:ha_sb_v_cont}
\end{figure}

\begin{figure}
\includegraphics[width=65mm,angle=270]{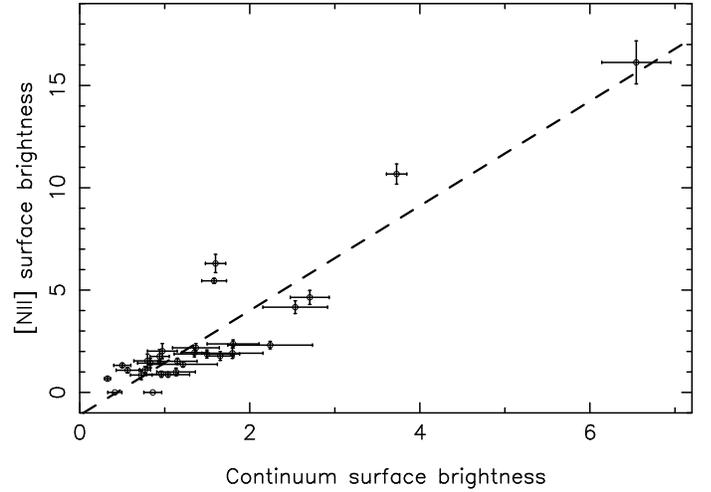}
\caption{[NII] line surface brightness vs red continuum surface
brightness
within star formation desert regions,
measured in units of 10$^{-20}$~W m$^{-2}$ sq. arcsec$^{-1}$
and 10$^{-20}$~W m$^{-2}$ \AA $^{-1}$ sq. arcsec$^{-1}$
respectively.}
\label{fig:nii_sb_v_cont}
\end{figure}

We next perform a simple test motivated by the possibility that the
diffuse line emission is causally linked to the old stellar population
present in the SFD regions, for example if the emission is due to 
p-AGB stars.  If this is the case, there should be a
correlation between the strength of the diffuse line emission and the
continuum surface brightness in the same regions, since the latter can 
most plausibly be assumed to be due to the old stellar population, in the
absence of ongoing or recent star formation.  Figures~\ref{fig:ha_sb_v_cont}
and \ref{fig:nii_sb_v_cont} illustrate a simple test of this possibility, 
where line surface brightness in W~m$^{-2}$ sq. arcsec$^{-1}$ in the H$\alpha$
and [NII] lines respectively are plotted against the red-light continuum flux
measured from the same spectra, in W m$^{-2}$ \AA $^{-1}$ sq. arcsec$^{-1}$.
Both figures show good correlations between line strength and old stellar
continuum, with Pearson correlation coefficients of 0.930 in both cases, and Spearman rank correlation coefficients of 0.836 and 0.846 for Figures~\ref{fig:ha_sb_v_cont} and \ref{fig:nii_sb_v_cont} respectively.  While such
correlation does not constitute proof, it is at least consistent with a link
between this line emission and the old stellar population.

\subsection{The galaxy-wide importance of diffuse line emission}

Given that we have identified near-ubiquitous diffuse line emission, it is of interest to investigate how significant this component is relative to the global emission-line activity of the galaxies in which it occurs.  For example, it would be useful to know what fraction of the total galaxy line flux, when imaged in a narrow-band filter, would be contributed by this diffuse component. To address this question, we have estimated the line fluxes integrated over these areas based on an estimated total H$\alpha +$ [NII] line flux.  To simulate narrow-band filter observations, this total comprised the sum of the H$\alpha$ line flux, without correction for absorption, and the [NII] 6584~\AA\ flux multiplied by a factor $1.0+(2.95)^{-1}$ to account for the weaker [NII] line at 6548~\AA, following the [NII] line ratios estimated by \citet{acke89}. The individual galaxy surface brightnesses were then integrated over the SFD areas of their galaxies in arcsec$^2$,
and converted into star formation rates in units of solar masses per year, using the conversion formula of \citet{kenn98}, giving a
mean of 0.0408 and a median of 0.0327~M$_{\odot}$~yr$^{-1}$ over the 15 galaxies.  To be clear, we only calculate these numbers for comparison purposes, and do not consider that ionisation by young stars is a likely explanation for this emission. The numbers presented have not been corrected for any internal extinction; using the correction preferred by \citet{kenn98} for bright spiral galaxies would increase these rates by a factor 2.8, but it is far from clear that these SFD regions contain significant amounts of dust.  

In any case, only one conclusion can be drawn from these equivalent star formation rates; the diffuse line emission we find in these barred galaxies is not a major contributor to the global line emission from these galaxies.  The total inferred SF rates can be compared with global rates derived from narrow-band imaging for 7 of the present sample in \citet{jame04}, which have a mean value of 2.02~M$_{\odot}$yr$^{-1}$.  So the diffuse emission from the SFD regions only comprises a few per cent of the total emission in these lines from these bright barred galaxies.  Future observations are planned to explore this conclusion further in different types of galaxies.

\section{Discussion}

In the analysis presented in this paper, we have used new spectroscopic 
data to investigate a region that is characteristic of a large fraction of strongly barred galaxies, which we have dubbed the star formation desert or SFD.  In earlier papers, we claimed that this region, which covers the radial range swept out by the bar, is devoid of line emission and hence star formation.  Using more sensitive long-slit spectroscopy, we have shown in the present paper that the first of these claims was incorrect, and in fact almost all of the SFD regions have detectable H$\alpha$ and [NII] emission.  However, this emission is diffuse and at a low surface brightness which accounts for its non-detection in previous narrow-band imaging studies.  The second claim appears to be correct, in that the emission line ratios appear to be inconsistent with those expected for star formation, with the [NII] line being relatively much stronger than is found in any star formation regions in this study, or indeed in the very extensive previous literature in this area.

We have put some additional constraints on the processes driving the diffuse line emission by the observation that the emission co-rotates at a rate that is closely coupled with the stellar disk, which argues against this being a very extended warm diffuse halo component, or shocked infalling or outflowing gas \citep{dopi95}.  The strength of the emission in both H$\alpha$ and [NII] correlates strongly with the strength of the underlying stellar continuum, which favours a link with p-AGB stars, which have been proposed in several previous studies as a source of diffuse line emission in normal, and particularly early-type galaxies with little or no star formation, e.g. \citet{stas08,sarz10,yan12,brem13,sing13}.  We note that AGN do not seem a likely source of this emission, since many of the 15 galaxies have no indications of AGN activity. In columns 3 and 4 of Table~\ref{tab:gals_obs}, 7 are listed as having AGN, LINER or composite nuclei, 4 are dominated by star formation and the remaining 4 have no specific nuclear classifications. In any case the emission is observed over several kpc which would argue against a single central ionising source.  Given the extent of the SFD regions, leakage of ionising photons from HII regions also seems to be ruled out.  However, there are some other intriguing possibilities for extended line emission which cannot be ruled out at present and should be mentioned for completeness.  Shock excitation of the gas is an interesting possibility, and we note that
\cite{gala99}, in a study of emission-line regions in M~31, found that supernova remnants which are 
believed to be dominated by shock excitation, showed [NII]/H$\alpha$ ratios in the range 0.48 - 1.2, mean 0.707, very similar to those of the SFD regions of the
present galaxy sample, after application of the  H$\alpha$ absorption correction. Another possible source of shocks is of course the bar itself, which motivates a further study of the strength of this diffuse emission as a function of angular separation from the bar axis.  Unfortunately this cannot be done with the present data set, due to only having single long-slit observations of each galaxy.

The recent study by \citet{sing13} provides a particularly interesting source of comparison for the present work, as it reached very similar overall conclusions, but by quite different methods. \citet{sing13} looked at 48 galaxies classified as showing LINER emission from within the CALIFA survey \citep{sanc12}, an integral field study of $\sim$600 galaxies in the low-redshift (0.005$<$ z $<$ 0.03) Universe. \citet{sing13} show convincingly that, while the LINER classifications are based on spectra extracted from a central region 3$^{\prime\prime}$ in diameter, the lines contributing to the LINER-type ratios fall off much more slowly than would be expected were they to be driven by a single nuclear source such as an AGN. Thus they strongly prefer a stellar explanation, such as p-AGB stars. It is also interesting to compare our results directly with two of the galaxies shown as examples in Figure 2 of \citet{sing13}.  This figure includes two star forming spiral galaxies, the first of which, NGC~4210, is also in the present study.  The emission-line spatial distribution of this galaxy, as found by \citet{sing13} very nicely matches the central peak/star formation desert/outer star formation ring pattern found here.  Interestingly, \citet{sing13} find a perfect trend for LINER-type line ratios in the central peak and star-formation line ratios in the outer disk, but they do not apparently detect any emission from the intermediate SFD region itself.  We do detect SFD emission in this galaxy; while not one of the strongest, [NII] emission is clearly detected in both SFD regions analysed, and H$\alpha$ emission is seen from one of them.  The other spiral galaxy shown in Fig.~2 of \citet{sing13} is also very interesting. This galaxy, UGC~11680, shows clear Seyfert-type emission from its nuclear region in the CALIFA data, with star-formation ratios in the spiral arms, and LINER-type ratios scattered across an extended region of the disk, in the interarm regions. This extended LINER emission is very similar to that found in the present paper. However, it is not clear whether UGC~11680 is directly comparable, as it may not be barred (the central bulge shows an extended morphology that is consistent with a weak bar, but the NED classification is Scd), and it is also undergoing a strong interaction with an early-type companion, which complicates the interpretation. 

\begin{figure}
\includegraphics[width=85mm,angle=0]{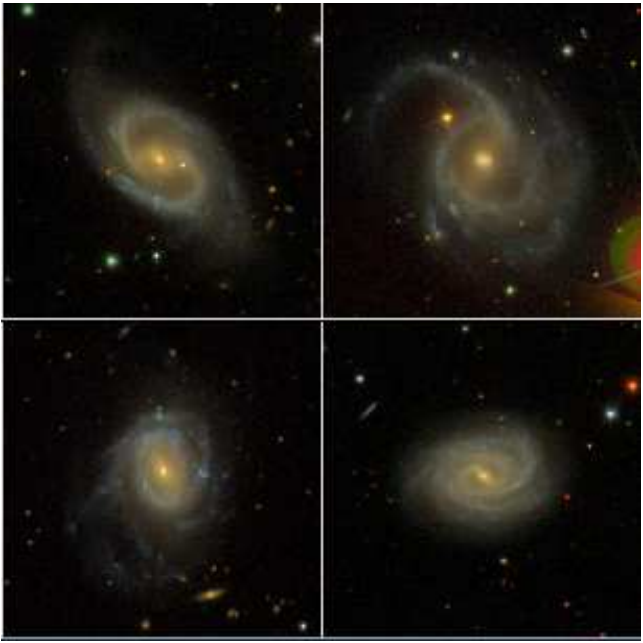}
\caption{SDSS images of (top left) NGC 2543 - SB(s)b; 
(top right) NGC 2595 - SAB(rs)c; 
(bottom left) NGC 3811 - SB(r)cd and 
(bottom right) NGC 4210 - SB(r)b.}
\label{fig:SDSS_four_galaxies}
\end{figure}

One other area requiring further study is the range of galaxies exhibiting this 
behaviour, as this study has quite explicitly made use of galaxies with carefully
defined properties.  The 15 galaxies studied are all clearly barred, with 10 having strong-bar SB classifications and the remaining 5 being SAB types, and they are generally
bright and massive spiral galaxies.  Most are quite early type spirals, with SBb 
being the most common type, and here it is relevant to mention the
bimodality of barred galaxies discussed by
\citet{nair10,hako14} (and confirmed by our analysis of galaxies from the H$\alpha$GS survey).
These authors separate out two types of bars: dynamically strong bars in early-type disk galaxies 
with dominant old, red stellar populations; and dynamically weaker bars in late-type, low-mass, gas-rich
spirals.
Analysis of the colour images produced from the multi-passband SDSS galaxy images illustrates this point well. As noted above, all but one of the present sample lie within the SDSS footprint, making such images available \citep{ahn14}, and an illustrative subset is shown in Fig.~\ref{fig:SDSS_four_galaxies}.  These images clearly show that the radial range swept out by the bar has red colours, indicating that an old stellar population is dominant, while the disk regions outside this region have blue colours indicative of ongoing star formation. This pattern is not seen for late-type barred galaxies of type SBd or similar, where the bar regions also show blue colours.
If this
bimodality in the effects of early- and late-type bars is real, the activity we are investigating in the present paper
appears to be linked with the early-type hosts, and to be most marked in SBb
types (range SBa - SBc approximately). This raises obvious questions about whether the activity
we find in the SFD regions of these galaxies is occurring in galaxies with later-type bars, and
indeed in galaxies generally.

The SDSS images also enable checks to be made of the environment of our galaxies.  Overall, they confirm that this is predominantly a sample of very isolated galaxies.  NGC~3185 is considered a member of the NGC~3190 group of galaxies, but it is well separated (11$^{\prime}$ - 17$^{\prime}$ projected separation) from the other three bright members of this group.  NGC~5806 is a member of the NGC~5846 group, but again is well separated (21$^{\prime}$) from the nearest group member. Finally, NGC~3367 is apparently close to the Leo group of galaxies, but is considered to be in the background due to a much higher redshift (3040~km~s$^{-1}$ c.f. 910~km~s$^{-1}$ for the group). Thus we conclude that none of the present sample has evidence for a close companion which could be causing significant tidal disturbance, or other interaction effects which might cause shock excitation of gas.

We note that galaxies showing the characteristic SFD identified here give a particularly
clean view of emission lines not associated with SF, in a region which 
is not accessible in unbarred galaxies, i.e. away from the nucleus, but 
within the central few kpc.  In unbarred spiral galaxies, this 
region is host to the dominant SF activity (in surface density, and as a fraction of the total
SF activity in these galaxies), making it impossible to detect low-level diffuse activity. 
The SFDs are an excellent environment for further study of these
anomalous line ratios, being a  clean stellar population with little contamination
from either AGN or SF activity.

\section{Conclusions}

The main conclusions of the present study are as follows:

\begin{itemize}
\item Barred galaxies, at least of earlier Hubble types, have strongly suppressed
star formation in the radial range swept out by their bars.
\item These extended `star formation desert' regions do however show line 
emission, with H$\alpha$ and/or the [NII] 6584~\AA\ line
being detected in virtually all of the barred galaxies in the present study.
\item The observed line ratios are not consistent with expectations for star formation regions, but have high
[NII]/H$\alpha$ ratios more characteristic of LINER-type emission. This conclusion holds regardless of the 
adopted correction for underlying stellar absorption at H$\alpha$.
\item The spatial distribution of line emission is also not consistent with expectations 
of star-forming  regions, being low surface brightness diffuse emission extended on scales of several kpc.
\item The diffuse emission lines show a strong rotation signal consistent with the kinematics of the stellar disk.
\item The strength of the diffuse components in both H$\alpha$ and [NII] emission is correlated with the red continuum from the local old stellar population.
\item The total emission integrated over the SFD region is typically only a few per cent of the line emission of the whole galaxy in which it sits.
\item Our preferred model is that this emission is powered by p-AGB stars from the old stellar population; AGN and leakage from HII regions do not seem plausible explanations, but shock processes cannot be ruled out and merit further investigation.
\end{itemize}

\section*{Acknowledgments}

The Isaac Newton Telescope is operated on the island of La Palma by the Isaac Newton Group in the Spanish Observatorio del Roque de los Muchachos of the Instituto de Astrof\'isica de Canarias.
We are happy to acknowledge support and assistance at the INT from Nicola Gentile, James McCormac and other ING staff. The referee, Artur A. Hakobyan, is thanked for a particularly thoughtful and helpful report, which included many suggestions that improved the content and clarity of this paper. 
This research has made use of the NASA/IPAC Extragalactic Database (NED) which is operated by the Jet Propulsion Laboratory, California Institute of Technology, under contract with the National Aeronautics and Space Administration.  
Funding for the SDSS has been provided by the Alfred P. Sloan Foundation, the Participating Institutions, the National Science Foundation, the U.S. Department of Energy, the National Aeronautics and Space Administration, the Japanese Monbukagakusho, the Max Planck Society, and the Higher Education Funding Council for England. The SDSS Web Site is http://www.sdss.org/.
The SDSS is managed by the Astrophysical Research Consortium for the Participating Institutions. The Participating Institutions are the American Museum of Natural History, Astrophysical Institute Potsdam, University of Basel, University of Cambridge, Case Western Reserve University, University of Chicago, Drexel University, Fermilab, the Institute for Advanced Study, the Japan Participation Group, Johns Hopkins University, the Joint Institute for Nuclear Astrophysics, the Kavli Institute for Particle Astrophysics and Cosmology, the Korean Scientist Group, the Chinese Academy of Sciences (LAMOST), Los Alamos National Laboratory, the Max-Planck-Institute for Astronomy (MPIA), the Max-Planck-Institute for Astrophysics (MPA), New Mexico State University, Ohio State University, University of Pittsburgh, University of Portsmouth, Princeton University, the United States Naval Observatory, and the University of Washington.


\begin{thebibliography}{}

\bibitem[\protect\citeauthoryear{Acker et al.}{1989}]{acke89} 
Acker A., K{\"o}ppen J., Samland M., Stenholm B., 1989, Msngr, 58, 44

\bibitem[\protect\citeauthoryear{Aguerri}{1999}]{ague99} Aguerri J.~A.~L., 1999, A\&A, 351, 43 

\bibitem[\protect\citeauthoryear{Ahn et al.}{2014}]{ahn14} 
Ahn C.~P., et al., 2014, ApJS, 211, 17

\bibitem[\protect\citeauthoryear{Baldwin, Phillips, 
\& Terlevich}{1981}]{bald81} Baldwin J.~A., Phillips M.~M., Terlevich R., 1981, PASP, 93, 5 

\bibitem[\protect\citeauthoryear{Bremer et 
al.}{2013}]{brem13} Bremer M., Scharw{\"a}chter J., Eckart A., Valencia-S.~M., Zuther J., Combes F., Garcia-Burillo S., Fischer S., 2013, A\&A, 558, A34 

\bibitem[\protect\citeauthoryear{Dopita 
\& Sutherland}{1995}]{dopi95} Dopita M.~A., Sutherland R.~S., 1995, ApJ, 455, 468

\bibitem[\protect\citeauthoryear{Ellison et 
al.}{2011}]{elli11} Ellison S.~L., Nair P., Patton D.~R., 
Scudder J.~M., Mendel J.~T., Simard L., 2011, MNRAS, 416, 2182 

\bibitem[\protect\citeauthoryear{Galarza, Walterbos, 
\& Braun}{1999}]{gala99} Galarza V.~C., Walterbos R.~A.~M., Braun R., 1999, AJ, 118, 277

\bibitem[\protect\citeauthoryear{Hakobyan et 
al.}{2014}]{hako14} Hakobyan A.~A., et al., 2014, MNRAS, 444, 
2428 

\bibitem[\protect\citeauthoryear{Hawarden et 
al.}{1986}]{hawa86} Hawarden T.~G., Mountain C.~M., Leggett 
S.~K., Puxley P.~J., 1986, MNRAS, 221, 41P 

\bibitem[\protect\citeauthoryear{Heckman}{1980}]{heck80} Heckman T.~M., 1980, A\&A, 87, 152 
 
\bibitem[\protect\citeauthoryear{Higdon, Buta, 
\& Purcell}{1998}]{higd98} Higdon J.~L., Buta R.~J., Purcell G.~B., 1998, AJ, 115, 80

\bibitem[\protect\citeauthoryear{Ho, Filippenko, 
\& Sargent}{1993}]{ho93} Ho L.~C., Filippenko A.~V., Sargent W.~L.~W., 1993, ApJ, 417, 63

\bibitem[\protect\citeauthoryear{Huang et 
al.}{1996}]{huan96} Huang J.~H., Gu Q.~S., Su H.~J., Hawarden T.~G., Liao X.~H., Wu G.~X., 1996, A\&A, 313, 13 


\bibitem[\protect\citeauthoryear{James et al.}{2004}]{jame04} 
James, P.~A. et al., 2004, A\&A, 414, 23

\bibitem[\protect\citeauthoryear{James et al.}{2009}]{jame09} 
James P.~A., Bretherton C.~F. \& Knapen J.~H., 2009, A\&A, 501, 207

\bibitem[\protect\citeauthoryear{Kennicutt}{1998}]{kenn98} Kennicutt R.~C., 
1998, ARAA, 36, 1089 
 
\bibitem[\protect\citeauthoryear{Knapen et al.}{2000}]{knap00} 
Knapen J.~H., Shlosman I. \& Peletier R., 2000, ApJ, 529, 93
 
\bibitem[\protect\citeauthoryear{Kormendy et al.}{2004}]{korm04} 
Kormendy J., \& Kennicutt R.~C., 2004, ARAA, 42, 603
 
\bibitem[\protect\citeauthoryear{Marinova et al.}{2007}]{mari07} 
Marinova I., \& Jogee S., 2007, ApJ, 659, 1176 

\bibitem[\protect\citeauthoryear{Nair 
\& Abraham}{2010}]{nair10} Nair P.~B., Abraham R.~G., 2010, ApJ, 714, L260
 
\bibitem[\protect\citeauthoryear{Percival et al.}{2009}]{perc09} 
Percival S.~M., Salaris M., Cassisi S. \& Pietrinferni A.,2009, ApJ, 690, 427 
 

\bibitem[Planck Collaboration et al.(2013)]{plan13} Planck 
Collaboration, Ade, P.~A.~R., Aghanim, N., et al.\ 2013, arXiv:1303.5062

\bibitem[\protect\citeauthoryear{S{\'a}nchez et 
al.}{2012}]{sanc12} S{\'a}nchez S.~F., et al., 2012, A\&A, 538, AA8 

\bibitem[\protect\citeauthoryear{Sarzi et al.}{2010}]{sarz10} 
Sarzi M., et al., 2010, MNRAS, 402, 2187

\bibitem[\protect\citeauthoryear{Singh et al.}{2013}]{sing13} Singh R., et al., 2013, A\&A, 558, AA43 

\bibitem[\protect\citeauthoryear{Stasi{\'n}ska et 
al.}{2008}]{stas08} Stasi{\'n}ska G., et al., 2008, MNRAS, 
391, L29 

\bibitem[\protect\citeauthoryear{Thilker et 
al.}{2002}]{thil02} Thilker D.~A., Walterbos R.~A.~M., Braun 
R., Hoopes C.~G., 2002, AJ, 124, 3118

\bibitem[\protect\citeauthoryear{Tubbs}{1982}]{tubb82} Tubbs 
A.~D., 1982, ApJ, 255, 458 

\bibitem[\protect\citeauthoryear{Yan 
\& Blanton}{2012}]{yan12} Yan R., Blanton M.~R., 2012, ApJ, 747, 61

\end{thebibliography}

\label{lastpage}

\end{document}